%% file: sigir2026.tex
\documentclass[sigconf]{acmart}
\AtBeginDocument{%
  }

\copyrightyear{2026}
\acmYear{2026}
\setcopyright{cc}
\setcctype{by}
\acmConference[SIGIR '26]{Proceedings of the 49th International ACM SIGIR Conference on Research and Development in Information Retrieval}{July 20--24, 2026}{Melbourne, VIC, Australia}
\acmBooktitle{Proceedings of the 49th International ACM SIGIR Conference on Research and Development in Information Retrieval (SIGIR '26), July 20--24, 2026, Melbourne, VIC, Australia}
\acmDOI{10.1145/3805712.3808421}
\acmISBN{979-8-4007-2599-9/2026/07}

\usepackage{amsmath}
\usepackage{natbib}
\usepackage{svg}
\usepackage{graphicx}
\usepackage{booktabs}
\usepackage{multirow}
\usepackage{multicol}
\usepackage{makecell}
\usepackage{caption}
\usepackage{enumitem}
\usepackage{graphbox}

\begin{document}

\title{SIGMA: A Semantic-Grounded Instruction-Driven Generative Multi-Task Recommender at AliExpress}

\author{Yang Yu}
\authornote{Corresponding author.}
\email{yy456487@alibaba-inc.com}
\affiliation{
  \institution{Alibaba International Digital Commercial Group}
  \city{Hangzhou}
  \country{China}
}

\author{Lei Kou}
\email{koulei.kl@alibaba-inc.com}
\affiliation{
  \institution{Alibaba International Digital Commercial Group}
  \city{Hangzhou}
  \country{China}
}

\author{Huaikuan Yi}
\email{yihuaikuan.yhk@alibaba-inc.com}
\affiliation{
  \institution{Alibaba International Digital Commercial Group}
  \city{Hangzhou}
  \country{China}
}

\author{Bin Chen}
\email{cb483157@alibaba-inc.com}
\affiliation{
  \institution{Alibaba International Digital Commercial Group}
  \city{Hangzhou}
  \country{China}
}

\author{Yayu Cao}
\email{caoyayu.cyy@alibaba-inc.com}
\affiliation{
  \institution{Alibaba International Digital Commercial Group}
  \city{Hangzhou}
  \country{China}
}

\author{Lei Shen}
\email{kenny.sl@alibaba-inc.com}
\affiliation{
  \institution{Alibaba International Digital Commercial Group}
  \city{Hangzhou}
  \country{China}
}

\author{Chao Zhang}
\email{chao.zc@alibaba-inc.com}
\affiliation{
  \institution{Alibaba International Digital Commercial Group}
  \city{Hangzhou}
  \country{China}
}

\author{Bing Wang}
\authornotemark[1]
\email{lingfeng.wb@alibaba-inc.com}
\affiliation{
  \institution{Alibaba International Digital Commercial Group}
  \city{Hangzhou}
  \country{China}
}

\author{Xiaoyi Zeng}
\email{yuanhan@alibaba-inc.com}
\affiliation{
  \institution{Alibaba International Digital Commercial Group}
  \city{Hangzhou}
  \country{China}
}

\renewcommand{\shortauthors}{Yang Yu et al.}

\input{data/abstract}




\maketitle
\input{data/introduction}
\input{data/method}
\input{data/experiment}

\input{data/conclusion}


\bibliographystyle{ACM-Reference-Format}
\bibliography{main}

\end{document}

%% file: data/abstract.tex
\begin{abstract}
With the rapid evolution of Large Language Models (LLMs), generative recommendation is gradually reshaping the paradigm of recommender systems.
However, most existing methods remain confined to the interaction-driven next-item prediction paradigm, struggling to keep pace with the latest evolving trends or address the diverse recommendation tasks along with business-specific requirements in real-world scenarios.
To this end, we present \textbf{SIGMA}, a \textbf{S}emantic-Grounded \textbf{I}nstruction-Driven \textbf{G}enerative \textbf{M}ulti-Task Recommender deployed at \textbf{A}liExpress.
Specifically, we first ground item entities in a unified latent space capturing both general semantics and collaborative signals.
Building upon this, we introduce a hybrid item tokenization method for both precise modeling and efficient generation.
Moreover, we construct a large-scale multi-task supervised fine-tuning dataset empowering SIGMA to fulfill various recommendation demands via instruction-following.
Finally, we design a three-step item generation procedure integrated with an adaptive probabilistic fusion mechanism to calibrate the output distributions based on task-specific requirements for recommendation accuracy and diversity.
Extensive offline experiments and online A/B tests demonstrate the effectiveness of SIGMA across various real-world recommendation tasks.


\end{abstract}

%% file: data/introduction.tex
\section{Introduction}

\begin{figure*}[!t]
    \centering
    \includegraphics[width=0.99\textwidth]{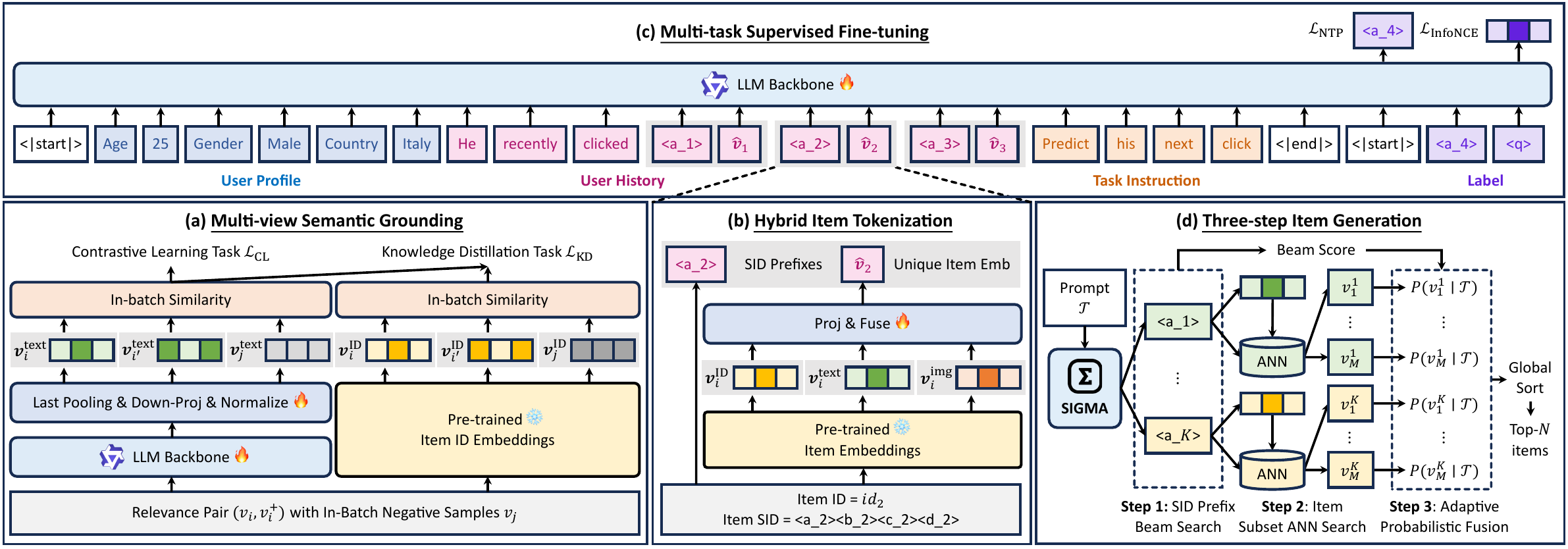}
    \caption{The overall framework of SIGMA.}
    \Description{The overall framework of SIGMA consists of four main components: multi-view semantic grounding, hybrid item tokenization, multi-task supervised fine-tuning, and three-step item generation.}
    \label{fig:framework}
\end{figure*}

Recently, the rapid advancement of Large Language Models (LLMs) has catalyzed a paradigm shift in recommender systems~\cite{wu2024survey,lin2025how,zhou2025onerectechnicalreport}.
Unlike traditional multi-stage cascaded pipelines, generative recommenders (GR) generate item identifiers in an end-to-end manner.
A growing body of studies has demonstrated its superior modeling capability, computational efficiency, and scalability~\cite{oneloc,liu2025generative,wei2024llmgr,onesearch,onesug}.

Despite recent progress, GR still faces several critical limitations.
First, the majority of existing works solely focus on the vanilla next-item recommendation task~\cite{onerec2025,lcrec,wang2024eager}.
They lack the versatility to accommodate diverse business-specific recommendation demands, such as providing personalized recommendations aligned with the upcoming festive theme or promoting specific items based on designated characteristics.
Besides, similar to traditional recommenders, mainstream GR methods remain heavily reliant on historical behavioral data~\cite{onerec2025,align3gr,liu2024sid}, resulting in an inherent delay when responding to dynamic trend shifts~\cite{mcdonald2023impatient,grbovic2018real}.
On the other hand, the strong semantic comprehension and extensive parametric knowledge of LLMs pave the way for a unified paradigm that consolidates heterogeneous recommendation demands via instruction-following~\cite{p5,lyu2024llm,llara}.
It also allows GR to rapidly address varied business requirements without waiting for the accumulation of user feedback data.

Given these factors, we propose \textbf{SIGMA}, a \textbf{S}emantic-Grounded \textbf{I}nstruction-Driven \textbf{G}enerative \textbf{M}ulti-Task Recommender deployed at \textbf{A}liExpress.
Specifically, to mitigate the domain shift and lack of collaborative signals for general LLMs~\cite{wang2024eager,qiao2024multi}, we first propose a multi-view alignment framework that grounds natural language, world knowledge, and item entities within a unified latent space capturing both semantic and collaborative relationships.
Unlike prevailing SID-based GR methods~\cite{align3gr,liu2024sid,hou2025sid}, we further develop a hybrid item tokenization strategy that combines SID prefixes with item-specific ID tokens to capture both structural and fine-grained characteristics of each item.
Furthermore, we curate a large-scale multi-task SFT dataset empowering SIGMA to fulfill diverse recommendation demands via instruction-following.
Finally, we introduce a three-step item generation procedure with an adaptive probabilistic fusion mechanism, which enables SIGMA to modulate its output distribution based on task-specific requirements for recommendation accuracy and diversity.
Our main contributions are as follows:
\begin{itemize}[leftmargin=*]
    \item We present SIGMA, a GR that can follow natural language instructions to satisfy diverse real-world recommendation demands.
    \item We propose a multi-view alignment framework which maps general semantics and platform-specific items into a unified embedding space modeling both semantic and collaborative relations.
    \item We design a novel hybrid item tokenization method along with a three-step item generation procedure and an adaptive probabilistic fusion mechanism for accurate and efficient GR.
    \item Extensive offline experiments and online A/B tests at AliExpress verify the versatility and efficacy of SIGMA across various tasks.
\end{itemize}

%% file: data/method.tex
\section{Methodology}

As illustrated in Fig.~\ref{fig:framework}, SIGMA consists of four main components: multi-view semantic grounding, hybrid item tokenization, multi-task supervised fine-tuning, and three-step item generation.

\subsection{Multi-view Semantic Grounding}\label{sec:emb}



To ground natural language, world knowledge, and item entities within a unified space, we curate a diverse set of relevance pairs covering both semantic and collaborative correlations:
\begin{itemize}[leftmargin=*]
    \item \textbf{Semantic Alignment}: We extract query-to-item pairs from the online search logs. Each query is linked with its high-click items.
    \item \textbf{Visual Alignment}: We leverage an online multi-modal model to pair items with the same-style visual features as relevance pairs.
    \item \textbf{Knowledge Alignment}: By integrating general LLMs with our in-site search system, we associate world knowledge (e.g., festivals, trending topics, seasonal shifts) with relevant themed items.
    \item \textbf{Collaborative Alignment}: We take items that frequently appear within the same interaction session as behaviorally correlated.
\end{itemize}

For each item, we first linearize its metadata (e.g., title, category, brand, etc.) into a JSON-formatted text sequence.
Then we employ a \emph{Contrastive Learning} task to fine-tune the general LLM.
Given a batch of relevance pairs $\{(v_i,v_i^+)\}_{i=1}^B$, we extract their semantic representations $\{\boldsymbol{v}_k^{\text{text}}\}_{k=1}^{2B}$ with the LLM, where $\boldsymbol{v}_{i}^{\text{text}}$ and $\boldsymbol{v}_{i+B}^{\text{text}}$ derive from ${v_i}$ and ${v_i^+}$, respectively.
We adopt the following InfoNCE Loss~\cite{infonce2018} with in-batch negatives for model optimization:
\begin{gather} 
    \mathcal{P}(\{\boldsymbol{v}_k^{\text{text}}\}_{k=1}^{2B},i,j)= \frac{e^{\cos(\boldsymbol{v}_i^{\text{text}},\boldsymbol{v}_{j}^{\text{text}})/\tau}}{\sum_{k=1,k\neq i}^{2B}e^{\cos(\boldsymbol{v}_i^{\text{text}},\boldsymbol{v}_k^{\text{text}})/\tau}}, \\
    \mathcal{L}_{\text{CL}}=-\frac{1}{2B}\sum_{i=1}^{2B}\log \mathcal{P}(\{\boldsymbol{v}_k^{\text{text}}\}_{k=1}^{2B},i,i^\prime),\label{infonce}
\end{gather}
where $i^\prime=(i+B)\ \text{mod}\ 2B$ and $\tau$ is a temperature hyper-parameter.

To bridge the gap between semantic and collaborative relevance, we further apply a \emph{Knowledge Distillation} task to transfer rich inter-item relationships captured by the item ID embeddings from the online ranking model.
For each item $v_i$, we take its ID embedding $\boldsymbol{{v}}_i^{\text{ID}}$ and finetune the LLM backbone to jointly optimize the following KL Divergence Loss, which aligns the in-batch similarity distributions derived from item semantic representations and ID embeddings:
\begin{equation}
    \mathcal{L}_{\text{KD}}=-\frac{1}{2B}\sum_{i,j=1}^{2B}\mathcal{P}(\{\boldsymbol{v}_k^{\text{ID}}\}_{k=1}^{2B},i,j)\log\frac{\mathcal{P}(\{\boldsymbol{v}_k^{\text{ID}}\}_{k=1}^{2B},i,j)}{\mathcal{P}(\{\boldsymbol{v}_k^{\text{text}}\}_{k=1}^{2B},i,j)}.
\end{equation}

\subsection{Hybrid Item Tokenization}\label{sec:hybrid}
To represent each item accurately and efficiently for SIGMA, we design a hybrid item tokenization method that combines Semantic ID (SID) prefixes with unique item IDs.
Specifically, we first employ RQ-VAE~\cite{tiger} to quantize each item semantic representation $\boldsymbol{v}_i^{\text{text}}$ into a sequence of hierarchical SID tokens $\{c_1, c_2, \dots, c_L\}$, where $L$ is the number of codebooks.
To mitigate the information loss during quantization, we further introduce a unique ID token for each item to supplement item-specific features.
Thus, each item $v_i$ is tokenized as $\{c_1, \dots, c_\ell, id_i\}$, where $\ell\in[1,L]$ is the length of the SID prefix.
After tokenization, each SID token is used to look up the LLM's token embedding table $\mathbf{E}$.
Meanwhile, the ID token $id_i$ serves as an index to retrieve various pre-trained item embeddings (e.g., ID embedding $\boldsymbol{v}_i^{\text{ID}}$, semantic embedding $\boldsymbol{v}_i^{\text{text}}$, visual embedding $\boldsymbol{v}_i^\text{img}$), which are integrated into a fused representation $\tilde{\boldsymbol{v}}_i$ via a learnable MLP.
The input embeddings of each item $v_i$ to the LLM backbone are formulated as $[\mathbf{E}(c_1), \dots, \mathbf{E}(c_\ell), \text{UpProj}(\text{MLP}(\boldsymbol{v}_i^{\text{ID}}, \boldsymbol{v}_i^{\text{text}}, \boldsymbol{v}_i^{\text{img}}))]$.

\subsection{Multi-task Supervised Fine-tuning}
To empower SIGMA with the capability to handle heterogeneous real-world recommendation demands, we construct a large-scale instruction-tuning dataset covering seven different tasks, which can be broadly categorized into four groups based on their objectives:

\begin{itemize}[leftmargin=*]
    \item \textbf{General Rec}: Focuses on fundamental personalized recommendations catering to user-specific interests (\texttt{JustForYou}).
    \item \textbf{Constrained Rec}: Targets precise and relevant recommendation subject to explicit constraints (\texttt{Query}, \texttt{Category}), requiring controlled generation within specific semantic subspaces.
    \item \textbf{Exploratory Rec}: Aims to surface niche items (\texttt{Longtail}) or identify latent user interests (\texttt{Discover}) by recalibrating output probabilities for low-frequency but high-potential candidates.
    \item \textbf{Common Sense Rec}: Leverages world knowledge to align recommendations with specific real-world contexts, such as seasonal shifts (\texttt{Season}) and holiday themes (\texttt{Holiday}).    
\end{itemize}

For each task, we construct extensive instruction-tuning samples consisting of the following parts: user's static profile $\mathcal{U}$ (e.g., age, gender, location), historical behaviors $\mathcal{H}$, a task-specific instruction $\mathcal{I}$, the SID prefix tokens of the target item $\{c_1,\dots,c_\ell\}$, and a special query token $q$ for target item ID prediction.
The LLM backbone is jointly optimized with two loss functions.
Specifically, the following Next Token Prediction Loss is calculated over the SID prefix tokens:
\begin{equation}
    \mathcal{L}_{\text{NTP}} = -\frac{1}{\ell}\sum_{t=1}^{\ell} \log P(c_t \mid \mathcal{U}, \mathcal{H}, \mathcal{I}, c_{<t}),
\end{equation}
For the target item ID, we adopt the following InfoNCE Loss~\cite{infonce2018} for efficient optimization across massive candidate items:
\begin{equation}\label{eq:infonce}
    \mathcal{L}_{\text{InfoNCE}} = -\log \frac{e^{\cos(\boldsymbol{h}, \tilde{\boldsymbol{v}}_i) / \tau}}{e^{\cos(\boldsymbol{h}, \tilde{\boldsymbol{v}}_{i}) / \tau} + \sum_{v_j\in\mathcal{N}_{\{c_1,\dots,c_\ell\}}} e^{\cos(\boldsymbol{h}, \tilde{\boldsymbol{v}}_{j}) / \tau}},
\end{equation}
where $\boldsymbol{h}$ is the hidden state of the query token and $\mathcal{N}_{\{c_1,\dots,c_\ell\}}$ is a set of randomly selected hard negative samples that share the same SID prefix with the target item.
$\tilde{\boldsymbol{v}}_i$ and $\tilde{\boldsymbol{v}}_{j}$ are the fused item representations of the target item and negative samples, respectively.

\subsection{Three-step Item Generation}
As illustrated in Fig.~\ref{fig:framework}(d), SIGMA adopts a three-step item generation procedure. Given a prompt $\mathcal{T}$, we first generate top-$K$ candidate SID prefixes and their beam scores with Beam Search as follows:
\begin{gather}
    \{c_1^k,\dots,c_\ell^k\}_{k=1}^K = \text{BeamSearch}(P(c_1,\dots,c_\ell\mid\mathcal{T});K),\\
    \phi_k = \sum_{t=1}^{\ell} \log P(c_t^k \mid \mathcal{T}, c^k_{<t}).
\end{gather}
Then we append the query token $q$ to each prefix and use its hidden state to retrieve the top-$M$ candidate items from the respective item subset via Approximate Nearest Neighbor (ANN) search as follows:
\begin{gather}
    \boldsymbol{h}_k = \text{SIGMA}(\mathcal{T},c_1^k,\dots,c_\ell^k,q),\\
    \{v_{i}^k\}_{i=1}^M=\text{ANN}(\boldsymbol{h}_k,\mathcal{V}_k;M)\approx\mathop{\raisebox{0.5ex}{$\text{arg top-}M$}}\limits_{v_j\in\mathcal{V}_k}\cos(\boldsymbol{h}_k,\tilde{\boldsymbol{v}}_j),
\end{gather}
where $\mathcal{V}_k$ denotes the set of items sharing the SID prefix $\{c_1^k,\dots,c_\ell^k\}$.

We further propose an Adaptive Probabilistic Fusion (APF) mechanism to yield the final recommendations based on task-specific requirements for precision and diversity.
Specifically, the generation probability of item $v_i=\{c_1^k,\dots,c_\ell^k,id_i\}$ given the prompt $\mathcal{T}$ is factorized into the following two components:
\begin{equation}
    P(v_i\mid\mathcal{T})=P(c_1^k,\dots,c_\ell^k\mid\mathcal{T})\times P(id_i\mid\mathcal{T},c_1^k,\dots,c_\ell^k).
\end{equation}
The first term corresponds to the beam score $\phi_k$ derived from generating SID prefixes, and the second term is formulated as follows:
\begin{equation}
P(id_i\mid\mathcal{T},c_1^k,\dots,c_\ell^k)=\frac{e^{\cos(\boldsymbol{h}_k,\tilde{\boldsymbol{v}}_i)\cdot\sigma(\{\phi_1,\dots,\phi_K\})/\tau}}{\sum_{v_j\in\mathcal{V}_k} e^{\cos(\boldsymbol{h}_k,\tilde{\boldsymbol{v}}_j)\cdot\sigma(\{\phi_1,\dots,\phi_K\})/\tau}},
\end{equation}
where $\sigma(\{\phi_1,\dots,\phi_K\})$ denotes the standard deviation of the beam scores across the top-$K$ SID prefixes.
For tasks with explicit constraints (e.g., \texttt{Query}, \texttt{Category}), there is typically a large gap between $\phi_1$ and $\phi_K$, leading to a larger $\sigma$ that sharpens the distribution toward high relevance items and ensures high precision.
Contrarily, tasks that require diversity (e.g., \texttt{JustForYou}, \texttt{Discover}) often have a smaller $\sigma$ which flattens the distribution, encouraging the exploration of non-top-ranked candidates.

Finally, all candidate items from different SID prefixes are ranked by their generation probabilities to yield top-$N$ recommendations.

%% file: data/experiment.tex
\section{Experiments}
\subsection{Experimental Setup}
\subsubsection{Datasets and Metrics}
We construct 150M relevance pairs for multi-view semantic grounding and 130M SFT samples based on the anonymized user behavior logs collected from AliExpress in 2026-01, with an additional 2M SFT samples from 2026-02 for evaluation.
All methods are evaluated in terms of average SID-level Hit Rate (HR@$K$) across all tasks.
For methods that directly generate item IDs, we map their predicted items to SIDs for a fair comparison.

\subsubsection{Baselines}
We compare SIGMA with the following baselines:
\begin{itemize}[leftmargin=*]
  \item Online Model: a Transformer-based generative model trained to predict users' next behavior in an autoregressive manner~\cite{gpsd}.
  \item GR (SID): An LLM-based GR which encodes items as SIDs derived from the semantic representations in Sec.~\ref{sec:emb} via RQ-VAE~\cite{tiger}.
  \item GR (ID): An LLM-based GR with ID-based item tokenization~\cite{urm}. It encodes each item with a fused embedding as detailed in Sec.~\ref{sec:hybrid}.
\end{itemize}
Both GR (SID) and GR (ID) are finetuned with the same SFT dataset.

\subsubsection{Implementation Details}
We use Qwen3-0.6B~\cite{qwen3} as the LLM backbone.
The dimensions of semantic representations and fused item embeddings are 64 and 128, respectively.
The hyper-parameter $\tau$ is set as 0.05.
We employ a 256$\times$4 codebook for SID generation.
The batch size for semantic grounding and SFT is 16,384 and 4,096, respectively.
We randomly select 100K batch-shared hard negative samples for Eq.~\eqref{eq:infonce}.
All models are trained using an AdamW~\cite{adamw} optimizer with a learning rate of 2e-5 and a 2,000-step linear warmup.

\subsection{Experimental Results}
\subsubsection{Performance Comparison}
\begin{table}[t]
\centering
\small
\caption{Performance of various methods. Bold denotes the best result, and the second best are underlined.}
\label{table:exp}
\resizebox{\columnwidth}{!}{
\begin{tabular}{lcccc}
\toprule
\textbf{Method} & \textbf{HR@1} & \textbf{HR@5} & \textbf{HR@10} & \textbf{HR@20} \\
\midrule

Online Model & 6.25\% & 15.16\% & 21.32\% & 30.06\%  \\
GR (SID) & 8.01\% & 16.17\% & 22.41\% & 26.41\% \\
GR (ID) & 7.31\% & 20.15\% & 28.98\% & 37.37\% \\
SIGMA (SID$_4$ID$\rightarrow$SID$_2$ID) & 8.91\% & 21.33\% & 28.43\% & 36.02\% \\
SIGMA (SID$_4$ID$\rightarrow$SID$_1$ID) & \underline{9.50\%} & \textbf{24.89\%} & \underline{33.37\%} & \underline{42.77\%}  \\
\textbf{SIGMA (SID$_1$ID $\rightarrow$ SID$_1$ID)} & \textbf{9.61\%} & \underline{24.73\%} & \textbf{33.76\%} & \textbf{43.05\%} \\
- Semantic Grounding & 7.80\% &	19.10\% & 25.85\% & 33.24\% \\
- APF (Top-$K$ 1st Prefix) & 9.31\% & 22.82\% & 30.18\% & 37.73\% \\
- Pre-trained Item Emb & 8.13\% & 21.82\% & 30.24\% & 39.11\% \\
- SID Prefix Hard Negative & 9.03\% & 23.79\% & 32.40\% & 41.34\%  \\
- \# Hard Negatives (10K) & 8.43\% & 22.87\% & 31.70\% & 40.86\% \\
\bottomrule
\end{tabular}
}
\end{table}
Table~\ref{table:exp} lists the performance of various baseline methods and SIGMA with different hybrid item tokenization settings, where SID$_x$ID$\rightarrow$SID$_y$ID denotes using $x$ SID prefix tokens in the prompt and $y$ SID prefix tokens for generation.
We first observe that SID$_4$ID$\rightarrow$SID$_2$ID yields inferior performance compared to SID$_4$ID$\rightarrow$SID$_1$ID.
This is because the generation accuracy of two SID prefix tokens is much lower than that of a single prefix token (63\% vs. 29\%), which limits the performance upper bound of the following item prediction.
However, SID$_4$ID$\rightarrow$SID$_1$ID brings marginal performance gains compared to SID$_1$ID$\rightarrow$SID$_1$ID, while substantially increasing sequence length and computational overhead.
Therefore, we adopt the SID$_1$ID$\rightarrow$SID$_1$ID setting for SIGMA, which significantly outperforms other baseline methods.

\subsubsection{Ablation Study}
We also validate the effectiveness of each component in SIGMA.
As shown in Table~\ref{table:exp}, ablating multi-view semantic grounding greatly impairs model performance, stressing the necessity to tackle domain shift and incorporate collaborative signals for general LLMs.
The APF mechanism also enhances the overall performance of SIGMA by tailoring output distributions to task-specific requirements.
Besides, integrating various pre-trained item embeddings and hard negative samples sharing the same SID prefix facilitates more accurate item modeling and generation.

\subsubsection{Model Scaling}
\begin{figure}[!t]
    \centering
    \includegraphics[width=0.83\columnwidth]{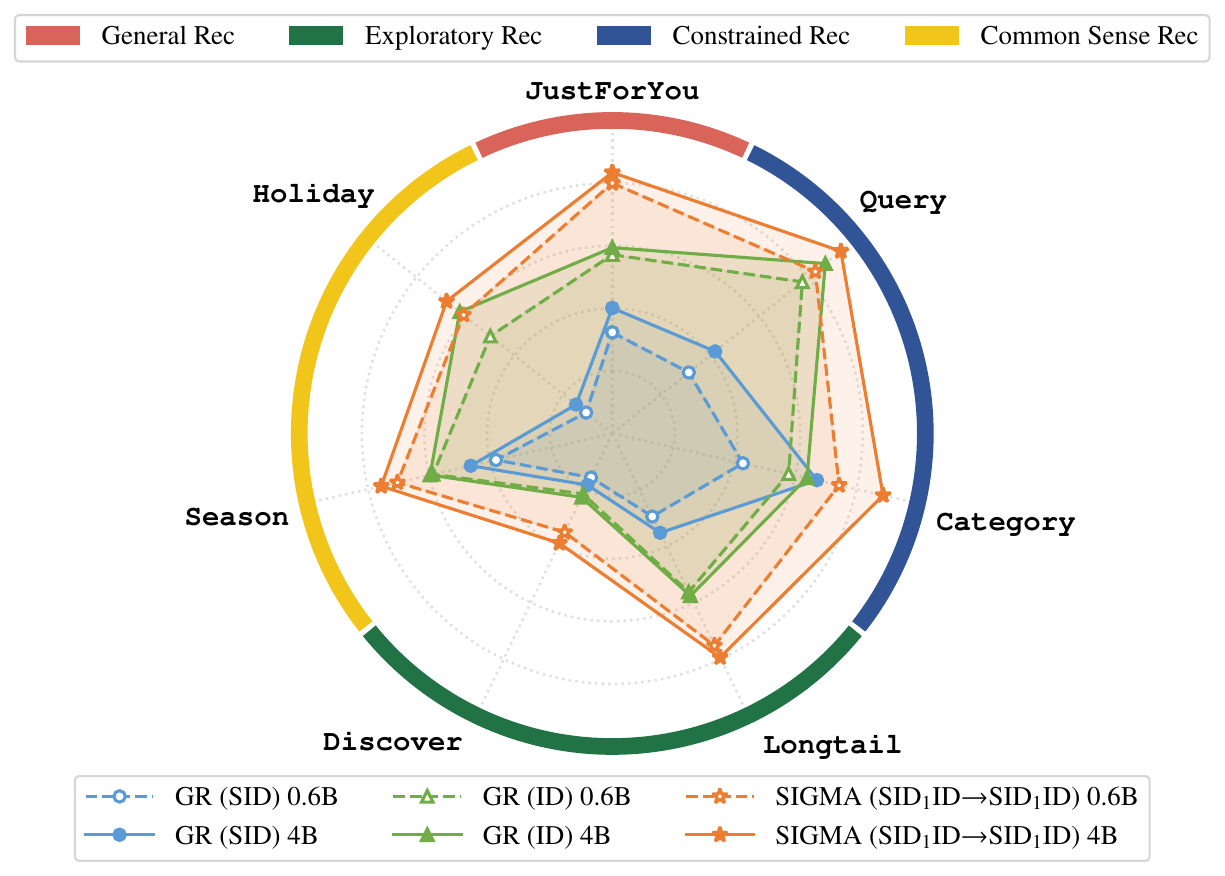}
    \caption{Performance variations of different methods on each task across with model scaling.}
    \Description{SIGMA achieves consistent performance improvement across all tasks with model scaling, especially in knowledge-intensive tasks and semantics-driven tasks.}
    \label{fig:radar}
\end{figure}
We further scale up the LLM backbone from Qwen3-0.6B to Qwen3-4B.
The performance variations of different methods on each task are illustrated in Fig.~\ref{fig:radar}.
The results show that SIGMA achieves consistent performance improvement across all tasks with model scaling, especially in knowledge-intensive tasks (\texttt{Holiday}) and semantics-driven tasks (\texttt{Query}, \texttt{Category}), which validates the effectiveness of employing larger LLM backbones with stronger semantic comprehension and richer world knowledge.

\subsection{Online Deployment and A/B Test}
\begin{figure}[!t]
    \centering
    \includegraphics[width=0.91\columnwidth]{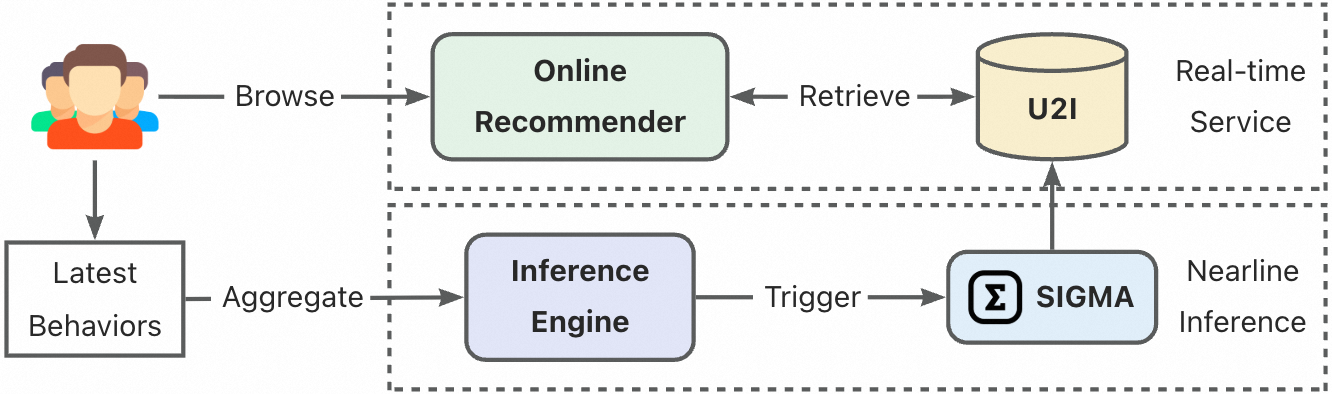}
    \caption{The online serving architecture for SIGMA.}
    \Description{We develop an online serving architecture consisting of nearline inference and real-time retrieval.}
    \label{fig:online}
\end{figure}
To meet the latency requirement of real-time service, we develop an online serving architecture consisting of \emph{nearline inference} and \emph{real-time retrieval} as illustrated in Fig.~\ref{fig:online}.
Users' latest behaviors are aggregated at minute-level intervals and trigger asynchronous incremental inference with past KV cache.
The resulting candidates are populated into a U2I index for real-time service.
We conduct 2-week online A/B tests for \texttt{JustForYou}, \texttt{Query}, \texttt{Season}, and \texttt{Holiday} tasks with 5\% traffic at AliExpress.
The results show that SIGMA-4B brings significant uplifts in key business metrics, including +2.80\% in Order Volume, +3.84\% in Conversion Rate, +7.84\% in GMV, and +2.47\% in Purchased Category Breadth, which verify its versatility and efficacy across various real-world recommendation tasks.

%% file: data/conclusion.tex
\section{Conclusion}
In this paper, we propose SIGMA, a versatile GR designed to address diverse recommendation demands through an instruction-following paradigm.
To deal with the domain shift and lack of collaborative signals for general LLMs, we first design a multi-view alignment framework that maps item entities into a unified latent space capturing both general semantics and collaborative signals.
Then we develop a hybrid item tokenization strategy which combines SID prefixes with unique item IDs for precise item modeling and efficient generation.
A large-scale instruction-tuning dataset is further curated for multi-task SFT.
Finally, we design a three-step item generation procedure integrated with an adaptive probabilistic fusion mechanism, which can tailor output distributions according to task-specific requirements for recommendation accuracy and diversity.
Extensive offline experiments and online A/B tests validate the effectiveness of SIGMA in diverse real-world industrial scenarios.